\documentclass[11pt]{article}
\pdfoutput=1

\usepackage{lineno,hyperref}
\usepackage{color}
\usepackage{amsfonts}
\usepackage{amsmath}
\usepackage{amssymb}
\usepackage{amsthm}
\usepackage{subeqnarray}
\usepackage{lscape,graphicx,url}
\usepackage{subcaption}
\usepackage{anysize}
\usepackage{proof}
\usepackage{fancyhdr}
\usepackage{latexsym}
\usepackage{array}
\usepackage{multirow}
\usepackage{textcomp}
\usepackage{optprog}
\usepackage{mathtools}
\usepackage{natbib}

\usepackage{bm}
\usepackage{array}

\usepackage{proof}
\usepackage{paralist}
\usepackage[dvipsnames]{xcolor}
\usepackage{placeins}
\usepackage{enumitem}
\usepackage{csquotes}
\usepackage{comment}
\usepackage[none]{hyphenat}

\newcolumntype{H}{>{\setbox0=\hbox\bgroup}c<{\egroup}@{}}

\title{Enhanced indexation using both equity assets and index options}

\author{Cristiano Arbex Valle$^1$ \and John E Beasley$^2$}

\date{}

\begin{document}

\maketitle

\begin{center} 
{\footnotesize
$^1$Departamento de Ci\^{e}ncia da Computa\c{c}\~{a}o, \\
 Universidade Federal de Minas Gerais, \\
 Belo Horizonte, MG 31270-010, Brasil \\
\{arbex\}@dcc.ufmg.br \\ \vspace{0.3cm}
$^2$Brunel University \\ Mathematical Sciences, UK \\ john.beasley@brunel.ac.uk
}
\end{center} 

\begin{abstract}

In this paper we consider how we can include index options in enhanced indexation. We present the concept of an \enquote{option strategy} which enables us to treat options as an artificial asset. An option strategy for a known set of options is a specified set of rules which detail how these options are to be traded (i.e.~bought, rolled over, sold) 
depending upon market conditions. 

We consider option strategies in the context of enhanced indexation, but we discuss how they have much wider applicability in terms of portfolio optimisation.

We use an enhanced indexation approach based on 
second-order stochastic dominance. We consider index options for the S\&P~500, using a dataset of daily stock prices over the period 2017-2025 
that has been
manually adjusted to account for survivorship bias. This dataset is made publicly available for use by future researchers.

Our computational results indicate that introducing option strategies in an enhanced indexation setting offers clear 
benefits in terms of improved out-of-sample performance. This applies whether we use equities or an exchange-traded fund as part of the enhanced indexation portfolio. 

\end{abstract}


\section{Introduction}

Enhanced indexation refers to the process of assembling a financial portfolio in an attempt to
outperform a specified marked index. Major market indices here include the  S\&P~500 in the USA and the FTSE~100 in the UK.

A common approach to addressing the problem of deciding an enhanced indexation portfolio is first to focus on the assets (equities, stocks)  in the index.
Then  to consider for these assets the
known  prices (and returns) in the immediate past, these being realisations of some underlying stochastic processes.  
Finally this past data is used in conjunction with some optimisation model to decide a portfolio that aims to outperform the index.

An implicit assumption here is  that any asset considered for inclusion in the enhanced indexation portfolio based on the immediate past will also exist into the immediate future.  For assets in major indices such as the S\&P~500 this assumption is valid. However this assumption is no longer true if we consider options.

Briefly here, since we discuss options in greater detail below, financial options exist relating to the market index value. For example a call option relating to the S\&P~500 would have a specified index value ($I$, say), at a specified time in the future  - so the option has a specified life before expiry.  If we choose to pay to buy this option now then we can, if we wish, sell it over its life.  At the end of its life, assuming we have not sold the option in the meantime, if the market index is above $I$ then we receive a cash amount depending upon the difference between the market index value and $I$. If the market index value is below $I$ then we receive nothing. 

But because an individual option has a specified life, and indeed can be sold over its life, the assumption referred to above, namely any asset considered for inclusion in the enhanced indexation portfolio based on the immediate past (the in-sample period)  will also exist into the immediate future (the out-of-sample period), is no longer true. This is especially so for the two S\&P~500 options considered in this paper which are: Monthly, expiring on the third Friday of every month and EOM, expiring on the last business day of each month. So, for example, if the in-sample period exceeds a month any option bought at the start of the period will have expired and we have the issue of what do we do with any cash generated?

The difficulty here is that we have a mismatch: equity assets have a \enquote{long life}, individual options have a \enquote{short life}. 
In this paper we propose an approach  based on \emph{\textbf{option strategies}} that enables us to overcome this mismatch, and 
hence consider both equity assets and options for inclusion in an enhanced indexation portfolio. 

The simple justification as to why we might
consider index options in enhanced indexation  is that index options
on major indices are widely available. As such including them in a portfolio seeking to outperform an index can, depending upon the decisions taken,
 contribute to achieving better outperformance than using equity assets alone.

An option strategy for a given option is a specified set of rules which detail how the option is to be traded  (bought, rolled over, sold) 
depending upon market conditions. Applying an option strategy to the immediate past means that we can evaluate it, 
so effectively it is equivalent to an asset with known performance. 
Because it is a strategy it has a \enquote{long life}, so can be chosen for inclusion in the enhanced indexation portfolio and applied in the future.
Each option strategy contributes one 
asset.  In other words we have converted a   \enquote{short life} option into a \enquote{long life}  option strategy asset. 

Option strategy  assets, together with equity assets, can be used in any optimisation approach for enhanced indexation. 
In this paper we make use of an approach based upon linear programming and  second-order  stochastic dominance. 

The structure of this paper is as follows. In Section~\ref{sec2} we review the relevant literature as to enhanced indexation and discuss 
previous work dealing with portfolio optimisation involving both equities and options. 
 We also state what we believe to be the contribution of the paper.
In Section~\ref{sec3} 
 we discuss the varying options that are available and how they are valued (priced), with special regard to index options for the S\&P~500.
We define what we mean by an option strategy, giving an 
 example S\&P~500 option strategy and 
illustrate how we can evaluate such a strategy.

In Section~\ref{sec4} we present our approach for enhanced indexation using both equity assets and options based upon linear programming and second-order stochastic dominance. In Section~\ref{sec5} we 
discuss the S\&P~500 asset dataset we used as well as  the option strategies considered. We give computational results illustrating the potential benefits of employing option strategies in an enhanced indexation setting. We consider investment in S\&P~500 equities and investment in a S\&P~500 exchange-traded fund both with, and without, option strategies.
In Section~\ref{sec6} we present a short discussion focusing on practical issues, option pricing and future work. Finally in Section~\ref{sec7} we give our conclusions.

\section{Literature review}
\label{sec2}
This paper deals with enhanced indexation making use of both equities and options.
\emph{\textbf{As far as we are aware there are no previous papers in the literature dealing with using options together with equities for enhanced indexation. Our paper is hence the first to deal with this problem.}} There are some papers dealing with using options together with equities for other portfolio optimisation problems and these are reviewed below.

\subsection{Enhanced indexation}
In this section we review the literature relating to enhanced indexation discussing papers that, in our view, are of more significance. 
We would
 mention here the recent work of~\cite{silva2024} who gave a comprehensive  literature review
which includes approaches for index tracking extended to  deal with enhanced indexation.

Early work dealing with enhanced indexation was given by~\cite{alexander05} who made use of cointegration
and~\cite{dose05} who clustered stocks based on a distance measure between two stocks defined using the difference between
their stock price histories (amongst other approaches).

 \cite{canakgoz09} presented a mixed-integer programming formulation that includes transaction costs, a cardinality constraint limiting the number of stocks that can
be purchased and a transaction cost constraint limiting the total transaction cost that can be incurred. Their approach makes use of the linear regression between portfolio returns and index returns. 
\cite{li2011} proposed a bi-objective formulation, minimising a tracking error function related to returns below the index, whilst maximising  portfolio return 
and limiting transaction costs.
\cite{meade2011} presented a paper that makes use of a modified Sortino ratio to produce portfolios of specified cardinality, but with an upper limit on
the holding in any particular asset. 

\sloppy \cite{guastaroba12} presented a mixed-integer programming formulation that includes transaction costs, 
a cardinality constraint and a transaction cost constraint.
Their objective is to minimise the absolute deviation between a scaled increased index value and the portfolio value. 
\cite{mezali2013} presented a mixed-integer programming formulation that includes transaction costs, 
a cardinality constraint and a transaction cost constraint.
Their objective is based on quantile regression. 

\cite{roman2013} made use of second-order stochastic dominance (SSD), aiming for a portfolio that is second-order 
stochastic dominant with regard to the index. Their solution procedure involves the use of cutting planes 
as given by~\cite{fabian2011a,fabian2011b}
who introduced a cutting plane reformulation of~\cite{roman2006}.
\cite{valle14}  
presented a three-stage  mixed-integer programming formulation that includes transaction costs, 
a cardinality constraint and a transaction cost constraint.
The first two stages relate to a regression of portfolio return against time, whilst the third stage relates to minimising transaction cost.

\cite{acosta2015} presented an approach based on optimising cointegration between the portfolio chosen and 
the benchmark index. They made use of genetic algorithms. For enhanced index tracking they artificially increased the benchmark index. 
\cite{bruni2015} proposed a linear bi-objective optimisation approach that maximises average
excess return and minimises the
maximum underperformance with respect to the index.
\cite{filippi2016} presented a bi-objective mixed-integer programming formulation balancing excess return against 
the absolute deviation between a scaled index value and the portfolio value.

\cite{guastaroba16}  presented a linear programming formulation based on the Omega ratio due to~\cite{omega}. This ratio involves 
setting a  predetermined threshold and then 
partitioning portfolio returns into those below the threshold (losses) and those above the threshold (gains). The Omega ratio is
defined as the probability weighted gains divided by the probability
weighted losses. 
\cite{bruni2017} proposed
 a criterion called ``cumulative zero-order stochastic $\epsilon$-dominance'' (CZS$\epsilon$D). Zero-order SD happens when all returns 
 from a given portfolio are superior to all returns from an alternative portfolio. The authors attempt to minimise underperformance by 
 adding an exponential number of constraints related to the  CZS$\epsilon$D criterion, where $\epsilon$ is the maximum 
 underperformance allowed. 

\cite{clark2019} presented an approach based upon reweighting the assets in the benchmark index. Their approach makes 
use of the concept of cumulative utility area ratio, which measures the incremental utility of one
asset compared to another.
\cite{guastaroba20} proposed  bi-criteria optimisation models using  a weighted 
combination of multiple conditional value-at-risk  (CVaR) measures.
\cite{li2021} presented a nonlinear enhanced indexation model with  an explicit
objective to track and outperform the naive diversification strategy. 

\cite{xu2022} make use of group-sparse optimisation, with dynamic parameter adjustment being based on either a hidden Markov
model or an artificial neural network.
\cite{allen2024} presented an approach based on robust
CVaR and group-sparse optimisation. 
\cite{dai2024} make use of a novel neural network together with a hidden Markov
model with Gaussian noise to capture regime switching. 
\cite{cesarone2025} presented an approach based upon ordered weighted average in conjunction with stochastic dominance. 
The weighted average is based on the sum of $k$ maximal CVaRs. 

With regard to the literature survey above two main themes can be seen:
\begin{compactitem}
\item One theme is the use of mixed-integer programming, where the integer variables are typically zero-one variables relating to the inclusion (or not) of an asset in the portfolio,  e.g.~\cite{canakgoz09, guastaroba12, mezali2013, valle14, filippi2016}.
\item Another theme, which typically only requires linear programming, makes use of stochastic dominance or CVaR,  e.g.~\cite{ roman2013, bruni2015, guastaroba16, bruni2017, guastaroba20, allen2024, cesarone2025}.
\end{compactitem}
\noindent As the reader may be aware solving a linear program is typically computationally less demanding  than solving a mixed-integer program. 

\subsection{Portfolio optimisation with equities and options}

In terms of papers dealing with using options together with equities for other portfolio optimisation problems a common assumption
 is that an option is held (unchanged) from portfolio creation until the end of the investment period. So options are only traded at portfolio 
 creation/rebalancing, no option trading is done over
  the investment period - in sharp contrast to the approach put forward in this paper.
  In addition some papers also draw on the framework established by~\cite{markowitz52}, balancing return against some risk measure.
  

Note that we only consider here work involving some optimisation approach to portfolio construction, excluding work 
such as~\cite{merton78, merton82, bookstaber84} that did not optimise portfolio composition. 

\cite{liang08} used a Markowitz framework with variance as the risk measure and presented approaches based on stochastic programming and stochastic 
control with dynamic programming.  \cite{scheuenstuhl08} used a Markowitz framework with variance as the risk measure and introduced 
shortfall constraints based upon lower 
partial moments. They also introduced monotony constraints to prevent speculation against the general market trend.

\cite{zylmer11} presented a robust optimisation model to maximise the 
worst-case portfolio return  if equity returns lie within their  uncertainty set. 
\cite{ashrafi21} presented a similar approach, but with a different uncertainty set.

\cite{davari16} used a multistage scenario tree with decisions (as to equities and options)  made at each node of the tree. Their objective is to maximise 
expected terminal wealth (after discounting) minus the present value of expected downside deviation (weighted at each scenario node) from a target wealth level. 
\cite{massar16,massar22} used a Markowitz framework with a target return, but varied the risk measure considering both variance and CVaR.

\cite{zhao18} extended the Markowitz framework, adopting a weighted objective balancing expected return 
against risk (variance)  and transaction cost.
\cite{khodamoradi20} used mean-absolute deviation  together with a cardinality constraint on the number of equities held
to find a portfolio involving both equities and options.
\cite{he23}  used the Markowitz framework adopting a weighted objective balancing expected return 
against  either variance or  CVaR. 

We would stress here that none of the papers reviewed in this section use options in a portfolio optimisation setting in as general a way as we propose in this paper, allowing trading of options before expiry in the investment period.

\subsection{Contribution}
\label{sec:ext}

Above we introduced the concept of an option strategy, defining for a given option a trading strategy which turns it from a  \enquote{short life} option into a \enquote{long life}  asset. 
This enables us to include options as  assets in enhanced indexation. As stated previously, as far as we are aware there are no previous papers in the literature dealing with using options together with equities for enhanced indexation. Our paper is hence the first to deal with this problem.

But we believe that the contribution of this paper reaches far beyond  enhanced indexation because of its potential impact. The work presented in this paper
 shows how options can be systematically and mathematically incorporated as assets into any portfolio optimisation problem. In the light of this we believe that the contribution of this paper is:
\begin{compactitem}
\item to introduce the concept of an option strategy

\item to, for the first time in the literature, include options as  assets in enhanced indexation

\item  to give computational results showing that introducing option strategies in an enhanced indexation setting offers clear benefits in terms of improved out-of-sample performance. This applies whether we use equities or an exchange-traded fund as part of the enhanced indexation portfolio.

\item to enable options (not just index options)  to be included  in many other situations. For example:
\begin{compactitem}
\item making use of  call and put options for individual  equity assets, e.g.~those that are traded on major companies in an index
\item considering other types of options, e.g.~futures
\item introducing options into any portfolio optimisation problem that involves equity assets, e.g.~Markowitz based approaches balancing return
against some risk measure
\item  trading options over time in an attempt to make money, but without involving equity assets
\end{compactitem}
\end{compactitem}

\section{Options}
\label{sec3}
In this section we discuss the varying options that are available and how they are valued (priced), with special regard to index options for the S\&P~500, which we use for our computational work below.
We then define what we mean by an option strategy, giving an 
 example S\&P~500 option strategy and 
illustrate how we can evaluate such a strategy.
 We go on to discuss how we can use option strategies in any optimisation based approach to enhanced indexation.
Finally we discuss the choice of an optimisation based enhanced indexation approach.

\subsection{Options and option pricing}
For  readers unfamiliar with options we give here the fundamentals of options and option pricing.

A call option on an underlying asset gives the holder of the option the right (but not the obligation) to buy the asset at
a fixed price, called the strike (or exercise) price. If this right can only be exercised at a specific (expiry)  time in the future then it is known as an European call option. If it can be exercised at any time up to expiry then it is known as an American call option.

A put option is the right to sell the asset, otherwise it has the same characteristics
as a call option
(so a strike price, expiry time, European or American).

Options create the potential for profit. For example suppose we buy an European call option and at expiry the 
underlying asset is priced in the market higher than the strike price. If using the call we buy the asset at the strike price and resell 
it at market price we have a potential profit (after accounting for transaction cost and the purchase cost of the call). However if at
 expiry the underlying asset is priced in the market lower than the strike price than the option is worthless.

A European put option is (potentially) profitable if at expiry the underlying asset is priced in the market below the strike price (as then 
we can buy the asset at market price and resell at strike price).

Even if an option can only be exercised at expiry it can be bought and sold at any time. So for example if we buy a call or put option now we can sell it
at any time between buying it and expiry. 

Clearly options can apply if we have a tangible asset, e.g.~a call option to buy one tonne of a particular metal. 
Similarly
in financial markets we can have options on specific equity assets, e.g. a call option to buy 100 shares of a particular stock. 

However options also apply if we have intangible assets, such as the S\&P~500 index. As we cannot \enquote{buy} any of the index an option such as this is cash settled. So the strike price relates to the level of the index. Considering a call option, if at expiry the index is below the strike price the option is worthless. However if at expiry the index is above the strike price then the option is cash settled, so the option holder receives a cash amount linearly related to the difference between the current index level and the strike price.

\sloppy Options can be priced (valued) using a model due to \cite{black73}, sometimes referred to as the Black-Scholes-Merton model based on~\cite{merton73}. Scholes and Merton were awarded the Nobel prize in Economics in 1997 for this work, Black was not eligible as he had died in 1995, see 
\url{https://www.nobelprize.org/prizes/economic-sciences/1997/press-release/}.  This pricing model, henceforth BS,  is as follows, let:
\begin{compactitem}
\item $t$ be the current time, where the option expires at time $T_e > t$, so the remaining
lifetime of the option is $L= T_e - t$, all times in years
\item $r$ be the annualised risk-free rate
\item $E$ be the exercise (strike) price of the option
\item $U_t$ be the current price  of the underlying asset (e.g.~the S\&P~500 if considering options on that index)
\item $\sigma$ be the  (implicit) volatility of the underlying asset 
\item $N(...)$  be the cumulative probability distribution function for the standard Normal distribution with mean zero and variance one
\end{compactitem}
Then the current price of a call option $P_{call,t}$ and the current price of a put option $P_{put,t}$ are given by:
\begin{equation}
P_{call,t} = U_{t}N(d_1) - E e^{-rL} N(d_2) 
\label{call}
\end{equation}
\begin{equation}
P_{put,t} = E e^{-rL} N(-d_2) - U_tN(-d_1)
\label{put}
\end{equation}
where
\begin{equation}
d_1  = \frac{\ln(U_t/E) +  (r + \sigma^2/2)L}{\sigma\sqrt{L}} 
\label{d1}
\end{equation}
\begin{equation}
d_2  = d_1 -\sigma\sqrt{L}
\label{d2}
\end{equation}
Pricing options using Equations~(\ref{call})-(\ref{d2}) is technically only applicable to European options where the asset is non-dividend paying. S\&P~500 options, which we deal with in this paper, are of this type. 

The distance 
between the exercise price and the current underlying price is called 
\emph{\it{moneyness}}. 
It is usually calculated with regards to a forward price
for the underlying asset 
(the S\&P~500  in our case). In this paper we calculate the forward price 
$F_t$ at time $t$ by assuming no dividends, 
a constant risk-free rate and a frictionless market, using:
\begin{equation}
F_t  = U_t e^{rL}
\label{eqForwardPrice}
\end{equation}
It is common to refer to options as at-the-money (ATM), in-the-money (ITM) or out-of-the-money (OTM). 

For a call option if $E=F_t$ it is ATM, if $E < F_t$ it is ITM, else it is OTM.
The logic behind ITM is that at expiry we can buy the underlying at $E$, less than the current forward price $F_t$, so giving a potential to make money.

For a put option  if $E=F_t$ it is ATM, if $E > F_t$ it is ITM, else it is OTM.
Table~\ref{table_money} summarises the situation.

Since (in practice) it is unlikely that $E$ exactly equals $F_t$ market participants refer to an option being ATM if $E \approx F_t$. 

\begin{table}[!ht]
\centering
{\small
\renewcommand{\tabcolsep}{1mm} 
\renewcommand{\arraystretch}{1.4} 
\begin{tabular}{|c|c|c|c|}
\hline
 Option type & ATM & ITM & OTM \\

\hline
Call & $E=F_t$ & $E <  F_t$ & $E  > F_t$ \\

Put & $E=F_t$ & $E >  F_t$ & $E  < F_t$ \\

\hline

\end{tabular}
}
\caption{At/in/out-of-the-money}
\label{table_money}
\end{table}

Numerically option moneyness (as a percentage) is defined as:
\begin{equation}
 100|F_t - E|/F_t
 \label{moneyeqn}
 \end{equation} 
So for example if $F_t= 100$  and $E = 103$ a put option is said to be $3\%$ ITM (ITM as $E > F_t$, see Table~\ref{table_money}).  If   $F_t = 103$  and $E = 100$ a put option is said to be $2.91\%$ OTM. 
The option strategies we present below make use of moneyness.

Academic readers interested in learning more about options are referred to a standard textbook such as~\cite{hull21}.

\subsection{Defining an option strategy}
\label{sec:str}
In enhanced indexation using equity assets we decide, for a particular equity asset, the cash amount to be invested in that asset, buy the asset and hold it into the future. So here once the amount to be invested in the asset is decided we have no more decisions to be made (technically until we rebalance our portfolio).

Using an option strategy for enhanced indexation the situation is much more complex. An option strategy for an option of a specific type is a set of rules that defines precisely how that option is to be traded as market conditions evolve. This complexity is a natural consequence of the fact that, as mentioned previously, an individual option has a short life.

For simplicity here we consider an option strategy involving just a single European option (e.g.~an index put option, such as considered below), but  the idea can extend to a set of options, an example of which is considered in our computational work later below.

In more detail we, as for equity assets, first decide for a particular option strategy the cash amount to be invested in that strategy. To define an option strategy we need rules that, at each and every time period, enable us to decide whether to:
\begin{compactitem}
\item Buy the option using cash.
\item Sell the option for cash.
\item Hold the option unchanged.
\item Rollover the option into another of the same type, but with a different lifetime and exercise price; so here we sell the option and immediately buy another of the same type.
\end{compactitem}
There are a number of assumptions that we need to make here:
\begin{compactitem}
\item If we buy the option using cash  all of the cash is used for option purchase.
\item If we rollover the option all the proceeds from sale are used to buy the new option of the same type.
\item Cash can be invested at the risk-free rate.
\item If the option has expired (so $L=0$) we exercise the option if it is profitable so to do.
\end{compactitem}
\noindent Collectively these assumptions mean that in any time period we are either holding cash or the option.
On a technical note here although we use phrases such as \enquote{buy the option} we mean buy a number of units of the same option. Since the sums of money involved are likely to be large we can ignore any issues relating to fractional option unit purchase.

Referring back to a point we made in the introduction to this paper it is clear that this option strategy has a \enquote{long life}. 
That is it can be applied consistently over time. By contrast a single option that we might purchase has only a \enquote{short life}.

So if we apply this option strategy as market conditions evolve we will produce a number of trades (or none if the strategy never leads to option purchase). Here we can be looking back historically (so in-sample) or moving forward in time (so out-of-sample). In each time period we may be holding an option or cash, so we have an \enquote{asset} value per period. From these we have  a set of per period returns, just as for equity assets we have a set of per period returns based on market prices.

\emph{\textbf{In other words an option  strategy is equivalent to an asset that can be treated just like an equity asset in any current approach for enhanced indexation}}.

\subsection{An example S\&P~500 option strategy}

In this section we give an example of a particular S\&P~500 index option strategy that is included in the computational results given below. First two facts about S\&P~500 options:
\begin{compactitem}
\item Exercise prices (the level of the S\&P~500 index at expiry)
are only integer multiples of 5.
\item In terms of expiry dates the two most liquid S\&P~500 options are: Monthly, expiring on the third Friday of every month and EOM, expiring on the last business day of each month.
\end{compactitem}

\noindent So consider a S\&P~500 put option where we have a $3\%$ OTM moneyness target. In more detail this strategy is:
\begin{compactitem}
\item Buy the option using cash if the index return over the past 30 days is $\leq-5\%$.
\item Sell the option for cash if the index return over the past 30 days is $>-5\%$.
\item Hold the option unchanged if not rolled over or sold.
\item Rollover the option into another of the same type if:
\begin{compactitem}
\item the lifetime of the option satisfies $L< 20$ days; or
\item the moneyness of the option has deviated significantly from the $3\%$ OTM target.
\end{compactitem}
\end{compactitem}
The economic logic behind this put option is that negative momentum (so a downward movement in the index) will continue.

Before we explain how we might rollover into another option of the same type it is clear here that this option strategy requires the user to have specified certain values. For example above the
 $3\%$ OTM moneyness target;
the index return over the past 30 days of 
$-5\%$, and the rollover when lifetime $L < 20$ days. By contrast if we are investing in equity assets once the amount to invest is decided no further decisions are needed (until rebalancing). 

Because of the necessity to decide numeric values in an option strategy there are (potentially) an unlimited number of option strategies that could be considered.

To illustrate how we rollover suppose the rollover rule above is active. So we are selling the current put option and buying another put option. For the purchased option we need to decide the 
\emph{\textbf{lifetime}} and the \emph{\textbf{exercise price}}. As the rollover rule involves $L < 20$ days to invoke sale of the current option then  for the purchased option we choose the next available expiry (monthly or EOM) 20 or more days in the future, giving the purchased option a lifetime of expiry time minus current time.

In terms of the exercise price $E$ for the purchased OTM option then as  we have a 
$3\%$ moneyness target and since for a OTM put option $E <F_t$ (see Table~\ref{table_money}) we have $100(F_t - E)/F_t = 3$ from Equation~(\ref{moneyeqn}), equivalently $E=0.97F_t$. Here $F_t$ is calculated from Equation~(\ref{eqForwardPrice}) using the lifetime as decided above for the purchased option.
As exercise prices are integer multiples of 5 we round $E$ to the nearest multiple of 5.

\subsection{ Example S\&P~500 option strategy evaluation}
\label{sec:eval}
As it can be deduced from  the previous section, evaluation of an option strategy over time using market data (in-sample or out-of-sample) is more complicated than evaluating holding an equity asset. This is because for an option strategy decisions whether to buy/sell/hold/rollover need to be made at each and every time period.

A further complicating factor is that, depending upon the option strategy and market conditions, we may on occasion
be holding no options. To illustrate this consider the example in Table~\ref{table2} for the option strategy given in the previous section.  Here for simplicity we ignore any rollover due to  deviation from the  3\% OTM moneyness target.

\begin{table}[!ht]
\centering
{\small
\renewcommand{\tabcolsep}{1mm} 
\renewcommand{\arraystretch}{1.4} 
\begin{tabular}{|l|c|c|c|c|c|c|}
\hline
Period & 1 & 2 & 3 & 4 & 5 & 6 \\

\hline
 S\&P~500 30 day return & -4.9\% & -5.1\% & -5.2\% & -5.2\% & -5.3\% & -4.8\% \\
\hline
Strategy decision         &             & Buy       & Hold    & Rollover  & Hold & Sell \\
Days to expiry & & 21 & 20 & 29 ($F_t=6379$) & 28 &  \\

Option price (sold) &  & &  & 80 &  & 110  \\
Option price (bought/held) &  & 90 & 95 & 75 & 77 &   \\
\hline
Exercise price &  &  &  & 6190 &  &    \\
Number of options held & & $\frac{1000u}{90}$ & $\frac{1000u}{90}$& 
$ \frac{80 \times1000u}{90\times 75}$
& 
$ \frac{80 \times1000u}{90\times 75}$ 
&  \\

Cash & 1000 &  &  &  &  & $ \frac{110\times 80 \times1000u}{90\times 75}$ \\

Valuation & 1000 & $1000u$ & $\frac{95(1000u)}{90}$ &
$ \frac{80 \times1000u}{90}$ &
$ \frac{77 \times 80 \times1000u}{90\times 75}$ 
 & $ \frac{110\times 80 \times1000u}{90\times 75}$  \\
\hline
Valuation (1dp, $u=1$) & 1000 & 1000 & 1055.6 & 888.9 & 912.6 & 1303.7   \\
Return (\%, 1dp) & - & 0 & 5.6 & -15.8 & 2.7 & 42.9 \\

\hline

\end{tabular}
}
\caption{Option strategy evaluation - example}
\label{table2}
\end{table}

In Table~\ref{table2} we have six time periods, where
all values from the exercise price row downward are calculated, all numeric values above that row are convenient values used for illustrative purposes.

Readers interested in a more detailed example of the use of the option strategy considered here with real world S\&P~500 data and option 
pricing using BS are referred to \url{https://github.com/cristianoarbex/ssdOptionsData}.

In the example in Table~\ref{table2}  we start in period 1 with 1000 in cash.  No options are bought in period 1 because the 30 day index return is not $\leq -5\%$. So at the start of period 2 we have $1000u$ to spend,  where $u$ is the  multiplier resulting from  depositing cash in an interest bearing account for one period.

In period 2  the 30 day index return is $\leq -5\%$ so we buy the option at 90 so in total $1000u/90$ options are bought, with 21 days to expiry.

In period 3 we hold the options bought (since neither the rollover nor sell option strategy conditions are satisfied)  with their new market value being 95 so the valuation of our holding of
$1000u/90$ options
is now $95(1000u)/90$.

In period 4 the options held in period 3 are now 19 days to expiry so we rollover, selling our current holding of $1000u/90$ options  for 80. For the new option the number of days to expiry is 29 (e.g.~EOM). Since, based on this new lifetime of 29 days, $F_t=6379$ the exercise price $E=0.97F_t$, which after
 after rounding to the nearest multiple of 5 is 6190. The new option costs 75 and we hold 
$(80{\times}1000u)/(90{\times} 75)$ units of that option.

In period 5 we hold the options bought (since neither the rollover nor sell option strategy conditions are satisfied)  with their new market value being 77 so the valuation of our holding of
$(80 {\times}1000u)/(90{\times} 75)$ options
is now $(77 {\times} 80 {\times}1000u)/(90{\times} 75)$.
In period 6 the 30 day index return is $> -5\%$ so we sell the option at 110.

The per period evaluation of this strategy (as formulae) is shown in the \enquote{Valuation} line in Table~\ref{table2}, being either the cash value or the value of the options held. This valuation evaluated numerically for $u=1$, so no interest is gained on cash, is given in the next to last row of the table, with
the return associated with this valuation being shown in the last row of the table.

We would comment here that, as the example above implicitly shows, we have no cost associated with option trading. In terms of options it is common to refer to both
transaction costs and liquidity penalties, the first of these being the normal cost of a trade and the second of these being the additional cost associated with an option trade where buyers/sellers of the option being traded are few.

It is clear however that if a value for the cost of a particular option trade can be produced 
then we can easily incorporate it into the evaluation of an option strategy. For example, referring to Table~\ref{table2}, if in period 2 the cost of buying one unit of an option is $Y$ then the effective cost of buying the option becomes $(90+Y)$, so the number of units bought becomes $1000u/(90+Y)$ with their valuation being $90 {\times} 1000u/(90+Y)$.  Adjusting the other values in Table~\ref{table2} for option trading cost can be done in a similar manner.

As an aside here transaction cost associated with trading equities is only incurred when we first buy the equity and at each rebalance. It can be accounted for in a similar fashion as described above.

Note here that we have excluded from consideration option purchase using debt finance (i.e.~taking on a debt to buy more of a particular option). Although this can be considered, it complicates the per period return calculation considerably if the option price changes so that the debt outweighs the value of the option holding (since then the strategy leads to a negative valuation).

\subsection{Option strategies and optimisation}
So consider the situation where we make use of a standard enhanced indexation optimisation approach, but include option strategies 
evaluated 
in-sample using past return behaviour as discussed above. 

It can arise that for some option strategies we never actively trade (buy/sell) an option when that strategy is evaluated in-sample. But, as illustrated by the example in Table~\ref{table2}, any investment allocated to that strategy will increase, e.g.~if $u > 1$, due to the risk-free rate. Logically this means that such an option strategy has no contribution to make in-sample, beyond accumulating cash at the risk-free rate.

To deal with this situation  we add investment in an appropriate interest bearing account (i.e.~at the risk-free rate) as a separate asset in the optimisation based solution approach given below and, at each rebalance, remove from consideration any option strategies that never actively trade an option when evaluated in-sample.
Since choice of an option strategy for inclusion in the portfolio to be held out-of-sample depends upon the 
behaviour of that strategy in-sample it seems logical that a strategy which never actively trades in-sample be excluded from consideration
for the out-of-sample portfolio.

 Optimising using in-sample data to decide the proportion of our wealth to be allocated to each asset (equity and option strategy) 
 we may find that no investment is made in  some option  strategies. 
 That is, given past history, these strategies do not seem to be worthwhile. However there may be one, or more, option strategies that are allocated some proportion of our wealth. Equally there may also be some equity assets that are allocated some proportion of our wealth. So based on in-sample data we have what would have been a good past investment using both equity assets and option strategies.
 
So now what do we do for the unknown future?
For any equity asset we simply buy the asset, using the proportion of wealth allocated to it, and hold it unchanged into the future.

For any option strategy we, using the proportion of wealth allocated to it, apply it into the future. 
However here the situation  with an option strategy is more complicated than for equity assets as the option strategy may, depending upon how it is constructed and future market conditions, never actually buy any options in the future. So the wealth allocated to it is left untouched.
Note here however that the
wealth allocated to it may increase at the risk-free rate, as in the example in Table~\ref{table2}.


\subsection{Choice of an optimisation based enhanced indexation approach}

In our literature review above we identified two main themes with regard to the literature and solution approaches for enhanced indexation, namely the use of mixed-integer programming and the use of linear programming.

In this paper we have chosen to make use of  linear programming. This is because, potentially at least, we might consider many different option strategies, certainly more than there are equity assets in the index. As such a computationally more effective (quicker) procedure, such as linear programming,  seems in our view to be the best choice.

We would strike a note of caution here that, because one might consider many different option strategies, there is a danger of in-sample over-fitting. By this we mean suppose that we produce millions of different option strategies. 
When we optimise to decide which ones to use based on historic data (so in-sample) the optimiser will naturally attempt to maximise its objective, choosing from the millions of strategies those that make best use of in-sample data.

However when we invest in these option strategies, i.e.~apply them out-of-sample into the future, we may find that by over-fitting in-sample 
we do not get good out-of-sample performance. In many respects this is not an issue when we  just use equity assets, 
since the number of such assets is typically limited by the number of equity assets in the index. In the results presented below we restricted ourselves to just a few option strategies to avoid any dangers of over-fitting.

\section{Second-order stochastic dominance}
\label{sec4}

In this section we present our approach for enhanced indexation using both equity assets and options based upon linear programming and second-order stochastic dominance (SSD). 

For readers unfamiliar with SSD it adopts a different approach from the work associated with the use of mixed-integer
programming discussed in Section~\ref{sec2} above. That work typically compares (enhanced) index return in period $t$ with portfolio return in period $t$  and optimises for portfolio composition. SSD breaks this chronological link between returns in the same time period and compares the $s$ smallest (enhanced) index returns with the $s$ smallest portfolios returns (irrespective as to when these $s$ smallest returns occurred) and optimises for portfolio composition.

For brevity here we assume that
the reader has some familiarity with SSD. 
Readers unfamiliar with this topic are referred to~\cite{fabian2011a, fabian2011b, roman2006}.

We would stress here that our option strategy approach can be applied to any existing approach for enhanced indexation. For computational reasons, as discussed above, we have chosen a linear programming based approach to  illustrate our approach. Let:
\begin{compactitem}
\item $[0,1,\ldots,T]$ be the historic time period over which we are looking back
\item $E$ be the set of equity assets 
\item $F$ be the set containing just the risk-free asset
\item $S$ be the set of option strategies, where we can subdivide $S$ into call only options $S_c$, put only options $S_p$ and mixed call/put options $S_{cp}$, where these subsets are disjoint and $S = S_c \cup S_p \cup S_{cp}$
\item $r_{it}$ be the return of asset $i \in E \cup F \cup S$ at time $t$, where for assets in $E \cup F$ these returns are computed from market data,
 for the option strategies 
 in $S$ these returns are computed by applying each strategy as discussed above
\item $r^I_{t}$ be the benchmark (index) return at time $t$
\item $\tau_s$ be the sum of the $s$ smallest index returns in $[ r^I_{1}, r^I_{2}, \ldots, r^I_{T}]$ divided by $T$
\end{compactitem}
Let $\mathcal{J}$ be a set, where each element of  
of $\mathcal{J}$  is itself a set of distinct time periods drawn from $[1,2,\ldots,T]$. If 
 $w_i$ is the proportion of our wealth that we invest in asset $i \in E \cup F \cup S$ the standard SSD approach to deciding these values is:
\begin{equation}
\mbox{maximise} ~ \mathcal{V}
\label{jebt4}
\end{equation}
subject to 
\begin{equation}
\mathcal{V}_s  \leq \frac{1}{T} \sum_{j \in \mathcal{J}} \sum_{i \in E \cup F \cup S} r_{ij} w_i - \tau_s~~~~\forall \mathcal{J} \subseteq \{1, ..., T\},~|\mathcal{J}| = s,~s=1,\ldots,T
\label{jebt5}
\end{equation}
\begin{equation}
 \mathcal{V}  \leq \mathcal{V}_s/ \kappa_s ~~~~s=1,\ldots,T
\label{jebt5fab}
\end{equation}
\begin{equation}
 \sum_{i \in E \cup F \cup S} w_i =1
\label{jebt6}
\end{equation}
\begin{equation}
 w_i \geq 0~~~~\forall i \in E \cup F \cup S
\label{jebt7}
\end{equation}
\begin{equation}
 \mathcal{V} \in\mathbb{R}
\label{jebt8a}
\end{equation}
\begin{equation}
 \mathcal{V}_s \in\mathbb{R}~~~s=1,\ldots,T
\label{jebt8}
\end{equation}

Here $\mathcal{V}_s$ is bounded from above by the difference between the $s$ smallest portfolio returns and the $s$ smallest index returns ($\tau_s$), each term divided 
by $T$, so a tail difference.
This is because on the right-hand side of Equation~(\ref{jebt5}) we consider all time period subsets  $\mathcal{J}$ of cardinality $s$ and subtract from
each of  them 
the constant term  $\tau_s$.
Equation~(\ref{jebt5}) is the standard SSD combinatorial definition of the tail differences, considering all subsets $\mathcal{J}$ of cardinality $s$.

In Equation~(\ref{jebt5fab}) $\kappa_s=1$ for unscaled SSD and 
$\kappa_s = (s/T)$ for scaled SSD. Equation~(\ref{jebt4}), in conjunction with Equation~(\ref{jebt5fab}),  maximises the minimum (scaled) tail difference.
Scaling, due to~\cite{fabian2011b}, gives more importance to the returns in the 
right tail of the distribution.  Since in this paper
we are seeking to outperform
the index emphasising the returns in the right tail seems appropriate, so we only use scaled SSD in our computational work below.

Equation~(\ref{jebt6}) ensures that all of our wealth is invested in assets. Equation~(\ref{jebt7}) is the non-negativity constraint 
(so no short-selling). Equation~(\ref{jebt8a}) ensures that  $\mathcal{V}$ can be positive or negative whilst
Equation~(\ref{jebt8}) ensures that the tail differences $\mathcal{V}_s$ can be positive or negative.

Equations~(\ref{jebt4})-(\ref{jebt8}) above is a portfolio choice optimisation program with explicit consideration of tails.
Although it has a combinatorial number of constraints, Equation~(\ref{jebt5}), these can be dealt with using a standard cutting plane procedure.
For completeness we give this cutting plane procedure in~\ref{append}.

 If the 
objective function has a non-negative optimal value then the associated portfolio  is second-order stochastic dominant with respect to the index.
Second-order stochastic dominance (if achieved) means that the sum of the $s$ smallest portfolio returns exceeds (or equals) the sum of the $s$ smallest index returns for all values of $s=1,\ldots,T$.

Equations~(\ref{jebt4})-(\ref{jebt8}) place no restrictions on the proportion of investment in equity assets $E$, the risk-free asset $F$,  or option strategies $S$. 
If $\omega(\Psi)$ and $\Omega(\Psi)$ are the lower and upper limits on the proportion of wealth invested in an asset subset $\Psi$ 
then we can add to this formulation the constraint:
\begin{equation}
\omega(\Psi)  \leq  \sum_{i \in \Psi} w_i \leq \Omega(\Psi) ~~~~\Psi=[E,F,S,S_c,S_p,S_{cp}]
\label{jebg}
\end{equation}
Clearly Equation~(\ref{jebg}) can be generalised to any subset of assets if the user has a particular requirement to 
group assets together and limit the exposure to that group in terms of overall investment.

\section{Computational experiments}
\label{sec5}

In this section we first discuss the asset dataset used, which is associated with equities in the S\&P~500 index. We then discuss the option strategies we used, and show computational results illustrating the benefits of employing option strategies in an enhanced indexation setting.
We  go on the discuss performance in different market regimes (bull/bear) and the potential impact of transaction cost.
 In our computational results we consider investment in S\&P~500 equities and investment in a S\&P~500 exchange-traded fund both with, and without, option strategies.

We used CPLEX Optimizer 22.1.0 (2023) as the linear programming solver, with default options. Our backtesting tool is developed in Python and all optimisation models are developed in C++. We ran all experiments on an Intel(R) Core(TM) i7-3770 CPU @3.90GHz with 8 cores, 8GB RAM and with Ubuntu 24 LTS as the operating system.
Detailed computation times are not given below but were typically very small, of the order of seconds per rebalance.

\subsection{Dataset}

The main dataset consists of daily stock prices from 17th March 2017 until 1st August 2025. This time period, over 8 years, includes the Covid pandemic, which had a significant effect on stock prices. Our data has been manually
adjusted to account for index composition. So on a given date only assets that were part of the S\&P~500 index at that time are available to be selected for investment, 
so we do not make use of any knowledge as to assets that might in the future become part of the S\&P~500 (or cease to be in the S\&P~500).
In order to define the scenarios required by SSD we used a lookback approach that included the most recent 201 daily prices, which then yield 200 in-sample returns.

The data also includes an exchange-traded fund (ETF), specifically SPY 
(\url{https://finance.yahoo.com/quote/SPY/}),
 which tracks the S\&P~500 index. SPY aims to replicate the performance of the S\&P~500 Net Total Return (NTR) index, which assumes reinvestment of dividends. Although SPY itself does not automatically reinvest dividends, its price reflects expectations about future payouts, and historical performance is closely aligned with the S\&P~500 NTR. In Section \ref{sec:etf}, we compare  buy-and-hold investment just in  SPY to an optimisation-based strategy where the asset universe is composed of both SPY and option strategies.

In this paper we deal with S\&P~500 options. We would anticipate that option prices as quoted in the market might differ from their theoretical BS price. However 
 for liquid (commonly traded) options the BS price is
\enquote{widely used}~\citep{brenner25}. In addition we lack access to historical data as to daily option prices over our time period (2017-2025). For these reasons we, in this paper, use BS to price the options we consider. 

Finally, we used the time series for the VIX index 
(\url{https://finance.yahoo.com/quote/%5EVIX/ })
as an approximation for  implied volatility for use in BS option pricing.
  We used the IRX (\url{https://finance.yahoo.com/quote/\%5EIRX})
 which represents the yield of the US 13-week Treasury bill,
  as a proxy for short-term risk-free investment and for pricing options with BS. The original IRX data is annualised, for use here we produce daily returns using the formula $r_{\text{daily}} = (1 + r^{\frac{1}{252}}) - 1$, where 252 represents the number of trading days in a year.

The data used in this paper is publicly available for use by other researchers at \url{https://github.com/cristianoarbex/ssdOptionsData}.

\subsection{Option strategies}

As an addition to the asset universe described above (S\&P~500 equities and SPY), we made use of
 12 option strategies based on three different policies. They are the following:

\begin{compactenum}
\item Buy/keep a put \{ATM, 3\% OTM\} option if the S\&P~500 index return over the past 30 days is $\leq \{-5\%, -10\%\}$. These four option strategies bet that negative momentum is likely to continue.

\item Buy/keep a call \{ATM, 3\% OTM\} option if the S\&P~500 index return over the past 45 days is $\leq \{-7\%, -15\%\}$. These four option strategies bet on a reversal of  longer term negative momentum.

\item Buy/keep a \{straddle, strangle\} when the annualised realised S\&P~500 volatility over the past \{20, 30\} days is below 8\%. 
These four option strategies bet on a \enquote{storm after a calm period}.
\end{compactenum}

\noindent For all of these strategies the rollover rule was to rollover if option lifetime satisfies $L< 20$ days or if
moneyness  has deviated significantly from the $3\%$ OTM target, so $100|0.97F_t - E|/0.97F_t \geq 3$. 
 Note especially here that with this lifetime ($L$) rollover rule 
 we never hold any option to expiry.
 
The economic logic behind these strategies is:
\begin{compactenum}
\item For  puts ATM and 3\% OTM, we have from 
Equation~(\ref{moneyeqn}) that the exercise prices at purchase are $E=F_t$ and $E=0.97F_t$. In this strategy we expect the index to continue falling, but if $U_t$ falls, so does $F_t$ (from Equation~(\ref{eqForwardPrice})). From Table~\ref{table_money} for a put if the (fixed) exercise price is above the forward price of the underlying it is ITM, so we would expect these puts to increase in value in the market. 

\item For  calls ATM and 3\% OTM, we have from 
Equation~(\ref{moneyeqn}) that the exercise prices at purchase are  $E=F_t$ and $E=1.03F_t$. For this strategy we expect the index to stop falling and rise, but if $U_t$ rises, so does $F_t$ (from Equation~(\ref{eqForwardPrice})). From Table~\ref{table_money} for a call if the (fixed) exercise price is less than the forward price of the underlying it is ITM, so we would expect these calls  to increase in value in the market. 

\item Straddles and strangles are non-directional trading strategies designed to profit from increases in volatility~\citep{hull21}.  A straddle buys an equal position in ATM calls and puts, with identical lifetimes and strike prices. It achieves a positive payoff when the underlying asset experiences sufficiently large price movements in either direction.
A strangle is similar, with the main difference being the purchase of a call and a put with different strike prices, typically with both options being OTM. 

\end{compactenum}

\subsection{Investment in equities}

In this section, we compare the performance of
\begin{compactitem}
\item using scaled SSD  to choose optimal portfolios from all S\&P~500 stocks (without options)
\item  using scaled SSD to choose optimal portfolios from all S\&P~500 stocks, but also including the 12 option strategies described above
\end{compactitem}
In both cases here we  include a risk-free investment asset, but make no use of SPY.

As mentioned above we used an in-sample period of 201 days. We conducted periodic rebalancing every 21 days (roughly one month in business days). 
To illustrate our approach our first in-sample period of 201 days runs from 17th March 2017 until 2nd January 2018. So using this in-sample period 
(with 200 return values for each asset) we choose a portfolio (using SSD) on 2nd January 2018, evaluate its performance out-of-sample for the next 20 business days 
(so from 2nd January 2018 to 31st January 2018, one day in this period being a US holiday) then repeat the process until the data is exhausted. In total this involved 91 out-of-sample periods, which we concatenate to have a single out-of-sample time series of returns. 

For simplicity we assume no transaction costs associated with trading equities or options  and no liquidity penalties when trading options. 
We impose no limits on the investment in assets, except that we impose a limit that at most 10\% of the portfolio can be held in the risk-free asset. 
As mentioned previously above  any option strategy whose in-sample evaluation is fully risk-free (so never trades an option) is excluded from the current rebalance.

\begin{figure}[!htb]

\centering
\includegraphics[width=1\textwidth]{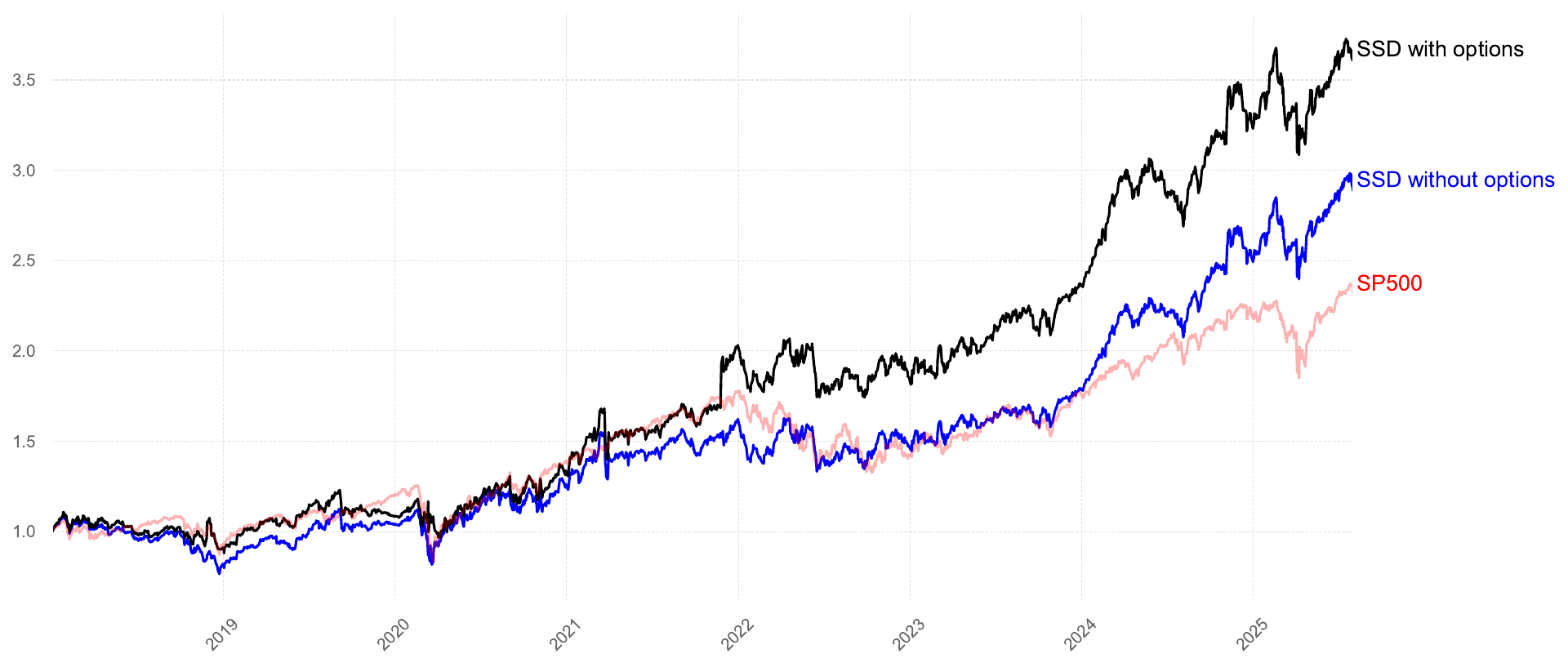}
  \caption{Scaled SSD with and without options, out-of-sample performance}
  \label{fig1}

\end{figure}

For numeric insight into these strategies, we show in Figure~\ref{fig1} a chart displaying their cumulative out-of-sample performance and in Table~\ref{table3} some selected comparative statistics. These are calculated from the out-of-sample returns of the two strategies, and correspondingly for the S\&P~500 index. 
It is clear from Figure~\ref{fig1} that SSD with options significantly outperforms SSD without options.

\begin{table}[!ht]

\centering
{\small
\renewcommand{\tabcolsep}{1mm} 
\renewcommand{\arraystretch}{1.4} 
\begin{tabular}{|l|rr|rr|rrr|}
\hline
\multicolumn{1}{|c|}{Strategy} & \multicolumn{1}{c}{FV} & \multicolumn{1}{c|}{CAGR} & \multicolumn{1}{c}{Sharpe} & \multicolumn{1}{c|}{Sortino} & \multicolumn{1}{c}{CVaR} & \multicolumn{1}{c}{Vol} & \multicolumn{1}{c|}{MDD}\\
\hline
SSD: equities and options     &     3.61 &    18.51 &     0.74 &     1.08 &     3.21 &    22.58 &    21.43\\
SSD: equities without options &     2.89 &    15.05 &     0.63 &     0.88 &     3.30 &    21.80 &    29.98\\

\hline
S\&P~500               &     2.31 &    11.74 &     0.52 &     0.72 &     3.04 &    19.96 &    33.92\\
\hline
\end{tabular}

\caption{Out-of-sample statistics comparing strategies based on scaled SSD with and without options}
\label{table3}
}
\end{table}

Let $Q$ be a series of $0,\ldots,T$ daily portfolio values, where $Q_t$ is the value of the given portfolio on day $t$. 
In Table~\ref{table3} \textbf{FV} stands for the final portfolio value, assuming a starting amount of \$1, and is calculated as $Q_{T}/Q_{0}$. \textbf{CAGR} stands for Capital Annualised Growth Rate and as a percentage is calculated as $100  \left( \left(\frac{Q_T}{Q_0}\right)^{\frac{1}{Y}} -1 \right)$, where $Y = T/252$ is an approximation for the number of years in the  period. \textbf{Sharpe} and \textbf{Sortino} are the annualised Sharpe and Sortino ratios respectively, where for their calculation we use the CBOE 10-year treasury notes (symbol TNX) as the risk-free rate.
 For  these four 
performance measures high values are better. 

\textbf{CVaR} represents the Conditional Value-at-Risk of the daily returns at confidence level 5\%, being a measure of tail risk. \textbf{Vol} represents the annualised sample standard deviation of the out-of-sample returns. \textbf{MDD} represents the maximum drawdown and is calculated as $\max \left(0, 100 \max_{0 \leq t < u \leq T} \frac{Q_t - Q_u}{Q_t} \right)$. All these three statistics are displayed as percentages, and for all of them low values are better.

In terms of performance, the plain SSD strategy (without options) compares favourably to the S\&P~500. On the other hand, the risk metrics are worse (CVaR and Vol), although the maximum drawdown was lower during the period. Despite that, the reward-risk metrics (Sharpe and Sortino) are higher than those of the S\&P~500.

The addition of the option strategies to the asset universe brings more diversification potential to SSD, including the potential to hedge against large drops in the market. With the exception of volatility, all statistics are better when comparing SSD with options to SSD without options. The hedge potential can be seen in the lower tail risk and maximum drawdown. 

Options are naturally much more volatile than regular stocks, so the increased volatility (22.58\% as opposed to 21.80\%) is unsurprising. However, note the increase of 17.5\% in the Sharpe ratio (0.74 as opposed to 0.63) and the increase of 22.7\% in the Sortino ratio (1.08 to 0.88). The denominator in the Sortino ratio considers only ``bad volatility'' (downside deviation), and the stronger increase means that most of the increased volatility is concentrated in the right tail (the ``good volatility''). 

\begin{figure}[!htb]
\centering
\includegraphics[width=1\textwidth]{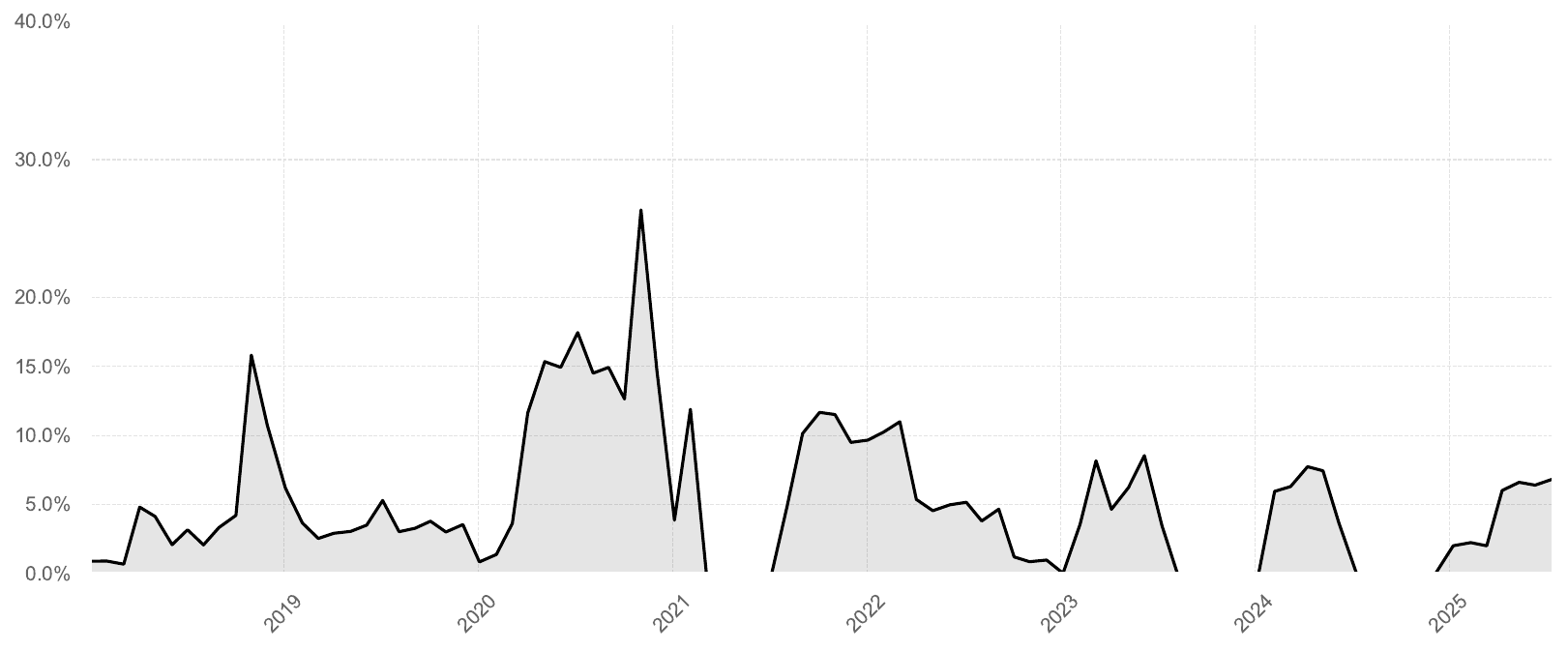}
  \caption{Accumulated SSD weights in all 12 option strategies over time}
  \label{fig2}
\end{figure}

SSD was allowed to freely choose any position in options, which can lead to very risky scenarios with high option exposure. Figure~\ref{fig2} shows the accumulated weight chosen by SSD in all option strategies together. From the figure we can see that most of the time the position in options was below 10\%, with occasional spikes, such as during the COVID era. We note that in rebalances where the COVID data was part of the in-sample period, we can see large positive returns in put options during a particular bad period, making them more likely to be chosen by SSD. Nevertheless, if necessary, exposure limits can easily be added to limit option exposure via use of Equation~(\ref{jebg}).

\subsection{Performance in different regimes}

\begin{figure}[htbp]
    \centering
    \begin{subfigure}{0.495\textwidth}
        \centering
        \includegraphics[width=\linewidth]{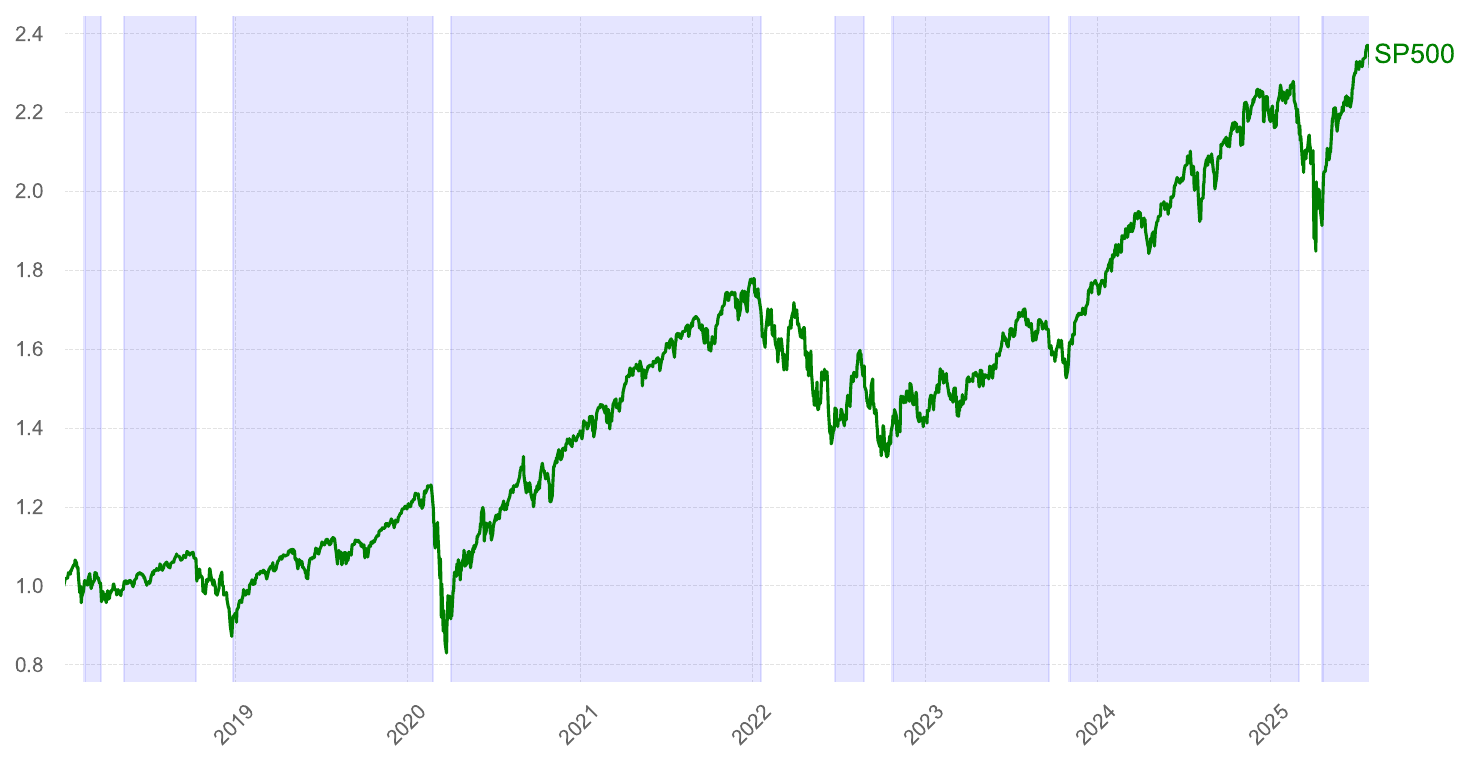}
        \caption{Periods classified as bull market}
    \end{subfigure}
    \hfill
    \begin{subfigure}{0.495\textwidth}
        \centering
        \includegraphics[width=\linewidth]{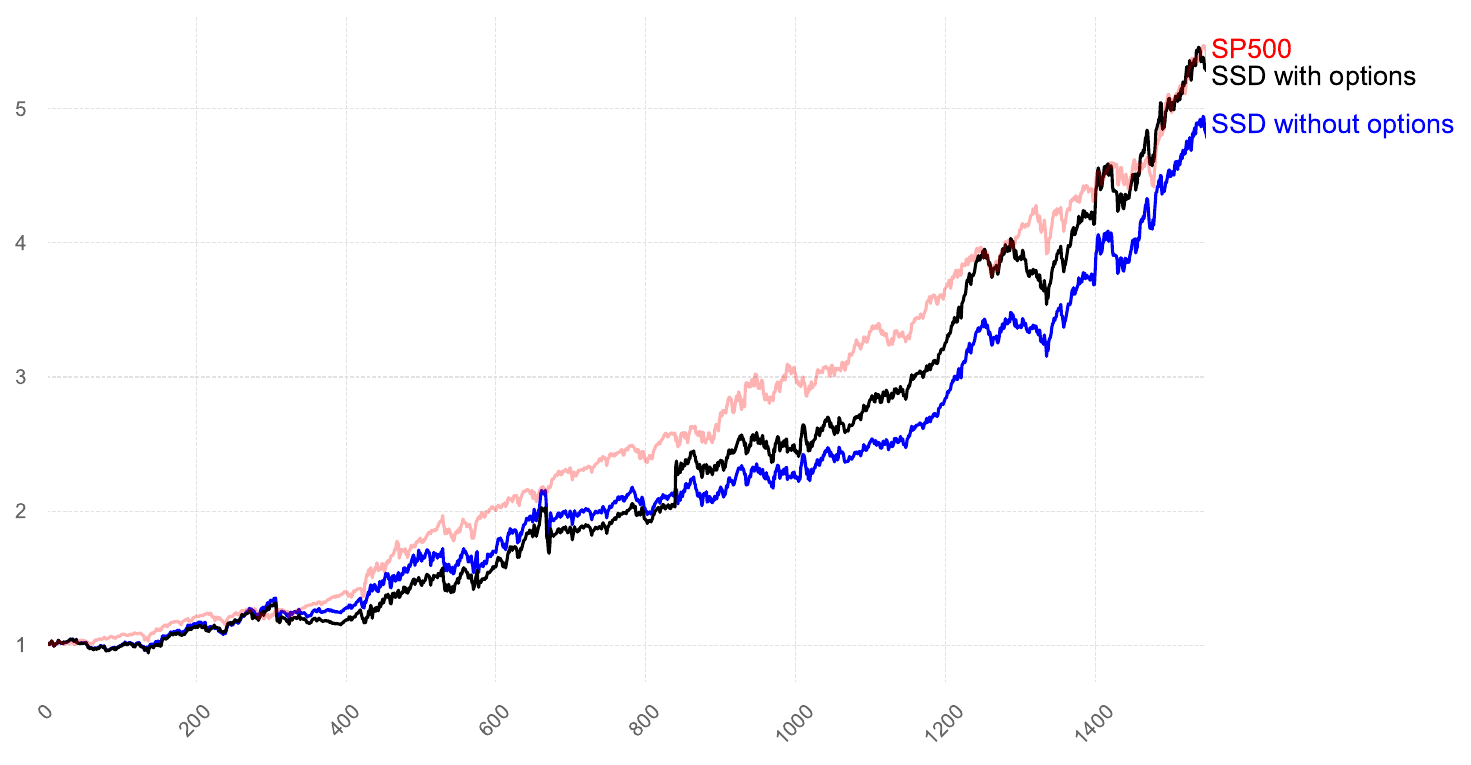}
        \caption{Out-of-sample performance during bull periods}
    \end{subfigure}

    \medskip

    \begin{subfigure}{0.495\textwidth}
        \centering
        \includegraphics[width=\linewidth]{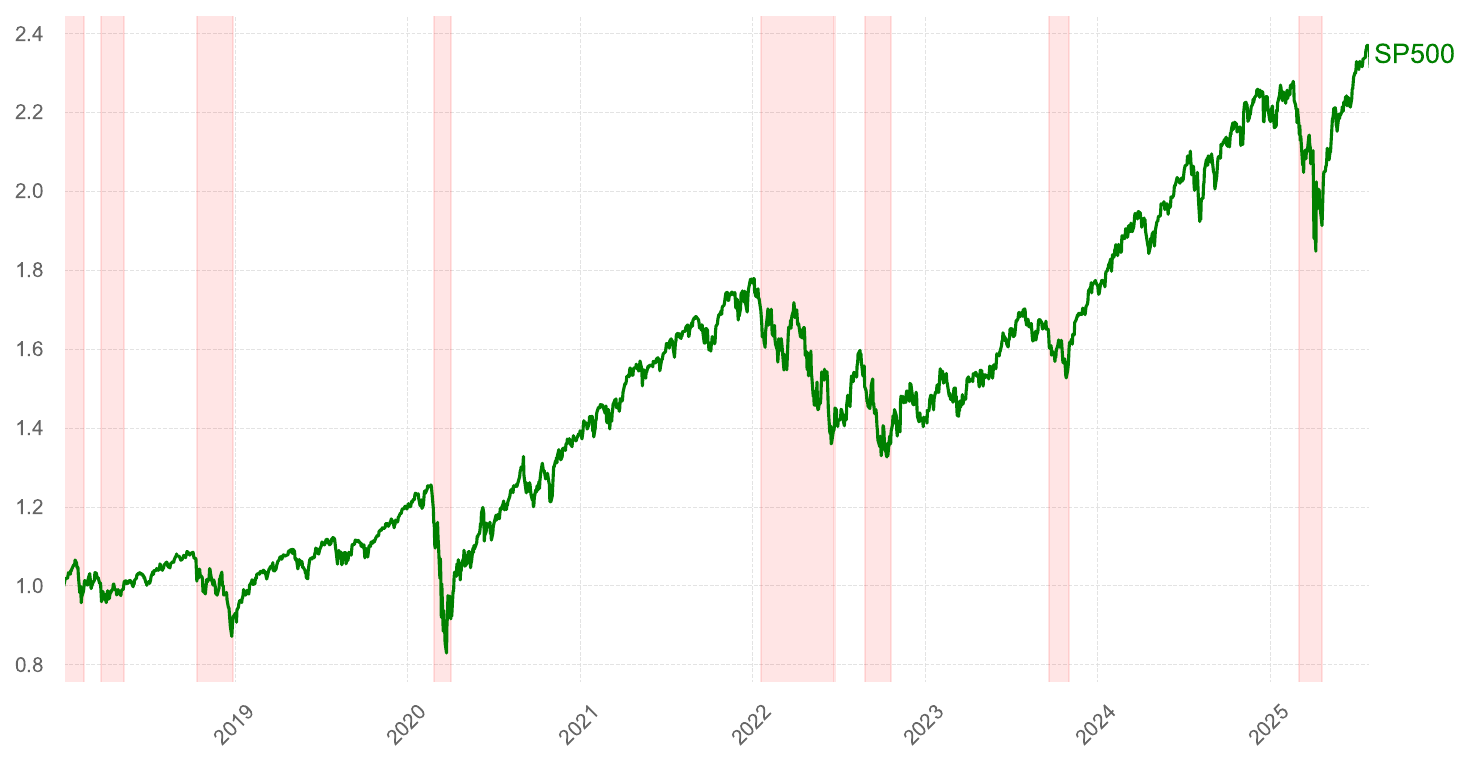}
        \caption{Periods classified as bear market}
    \end{subfigure}
    \hfill
    \begin{subfigure}{0.495\textwidth}
        \centering
        \includegraphics[width=\linewidth]{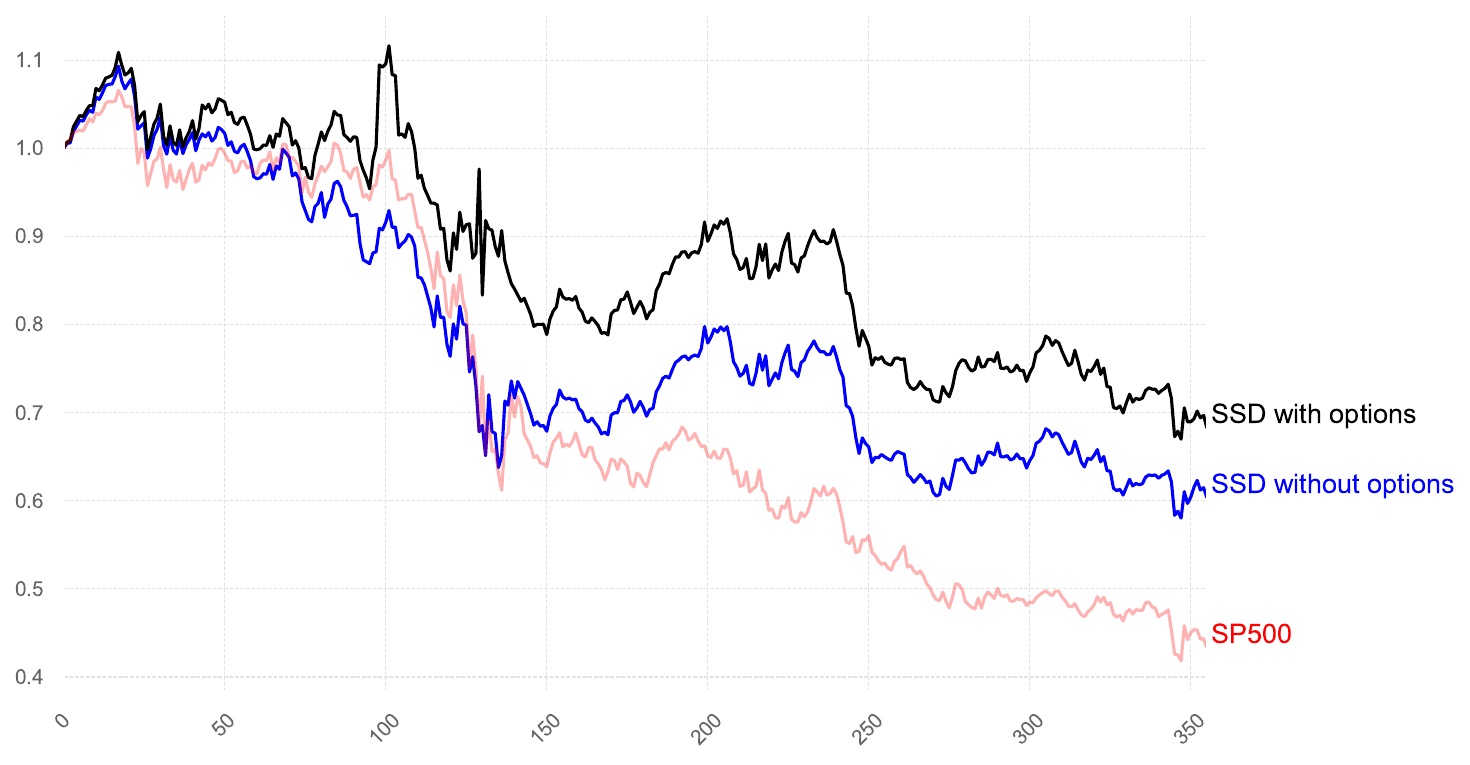}
        \caption{Out-of-sample performance during bear periods}
    \end{subfigure}

    \caption{Out-of-sample performance across different regimes}
    \label{figRegime}
\end{figure}

In this section, we evaluate the out-of-sample performance of both strategies and the S\&P~500  conditional on market regimes. 
We consider both bullish (prices rising) and bearish (prices falling) markets. 
We manually classified the S\&P~500 price history as either bull or bear according to the visually observed price trend. Figure~\ref{figRegime} includes four panels and shows both the visual regime classifications and the out-of-sample performance across each regime.

In panels (a) and (c), we highlight (using vertical shading) the periods that are manually classified as either bull or bear markets; each period is assigned to exactly one regime. Panels (b) and (d) present regime-conditional out-of-sample performance. We constructed each regime-specific return series by concatenating the strategies and index returns corresponding to the highlighted days into a set of continuous asset return time series.
Note here that panel (b) involves over 1400 days, panel (d) only around 350 days.

\begin{table}[!ht]
\centering
{\small
\renewcommand{\tabcolsep}{1mm} 
\renewcommand{\arraystretch}{1.4} 
\begin{tabular}{|l|l|rr|rr|rrr|}
\hline
\multicolumn{1}{|c|}{Regime} & \multicolumn{1}{c|}{Strategy} & \multicolumn{1}{c}{FV} & \multicolumn{1}{c|}{CAGR} & \multicolumn{1}{c}{Sharpe} & \multicolumn{1}{c|}{Sortino} & \multicolumn{1}{c}{CVaR} & \multicolumn{1}{c}{Vol} & \multicolumn{1}{c|}{MDD}\\
\hline
\multirow{3}{*}{Bull market} &   SSD: equities  with options    &     5.29 &    31.11 &     1.32 &     1.97 &     2.81 &    20.09 &    16.87\\
& SSD: equities without options &     4.78 &    28.97 &     1.27 &     1.82 &     2.82 &    19.42 &    16.87\\              
                             & S\&P~500               &     5.34 &    31.32 &     1.74 &     2.62 &     2.08 &    14.79 &     9.60\\
\hline
\multirow{3}{*}{Bear market} & SSD: equities  with options    &     0.68 &   -23.68 &    -0.81 &    -1.10 &     4.45 &    31.10 &    39.99\\
                              & SSD: equities without options &     0.60 &   -30.06 &    -1.15 &    -1.55 &     4.50 &    29.92 &    46.91\\
                             & S\&P~500               &     0.43 &   -44.70 &    -1.65 &    -2.16 &     5.15 &    34.12 &    60.75\\
\hline
\end{tabular}
}
\caption{Out-of-sample statistics across different market regimes}
\label{tableRegime}
\end{table}

Comparative regime-conditional statistics are shown in Table~\ref{tableRegime}. Unsurprisingly, during periods classified as bullish, the S\&P~500 showed strong performance and relatively low risk. The two SSD strategies had similar risk profiles, higher than the index, but the strategy with options achieved better returns, almost matching that of the S\&P~500. 
During periods classified as bearish, both strategies strongly outperformed the S\&P~500, both in terms of risk and return. SSD with options out-performed  SSD without options on all but one measure.

Overall, the addition of option strategies enhanced the performance of SSD both by increasing upside potential during ``good'' times and mitigating downside risk during ``bad'' times.

\subsection{Investment in SPY}
\label{sec:etf}

Passive investing through the purchase of ETFs and index funds has been increasing. In 1993  passive funds invested in US stocks managed \$23 billion of assets, 
some 0.44\% of the US stock market. By 2021 passive assets had risen to \$8.4 trillion, 
some 16\% of the stock market~\citep{jiang2025}.

\begin{figure}[!htb]

\centering
\includegraphics[width=1\textwidth]{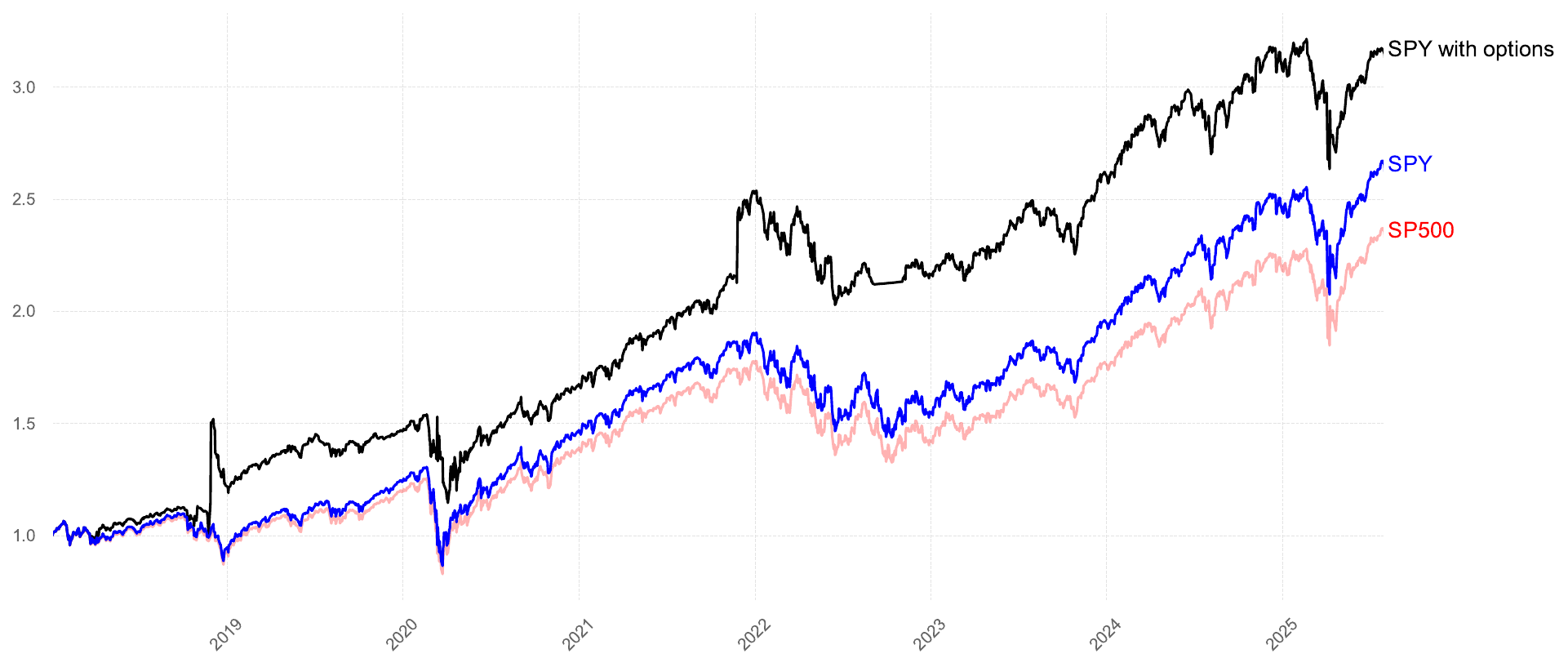}
  \caption{SPY and a scaled SSD strategy including SPY and options, out-of-sample performance}
  \label{fig3}

\end{figure}

In this section we evaluate whether it would be profitable to adopt a semi-passive strategy which uses the scaled SSD model with an
asset universe composed of just  the SPY ETF and the 12 option strategies. Figure~\ref{fig3} shows  out-of-sample cumulative returns, and Table~\ref{table4} displays selected out-of-sample statistics. Both follow the same format as Figure~\ref{fig1} and Table~\ref{table3}.
Again we see that using (scaled) SSD with options performs best.

\begin{table}[!ht]

\centering
{\small
\renewcommand{\tabcolsep}{1mm} 
\renewcommand{\arraystretch}{1.4} 
\begin{tabular}{|l|rr|rr|rrr|}
\hline
\multicolumn{1}{|c|}{Strategy} & \multicolumn{1}{c}{FV} & \multicolumn{1}{c|}{CAGR} & \multicolumn{1}{c}{Sharpe} & \multicolumn{1}{c|}{Sortino} & \multicolumn{1}{c}{CVaR} & \multicolumn{1}{c}{Vol} & \multicolumn{1}{c|}{MDD}\\
\hline
SSD: SPY and options &     3.12 &    16.25 &     0.65 &     1.11 &     2.83 &    22.67 &    25.57\\
SPY              &     2.61 &    13.51 &     0.60 &     0.84 &     3.02 &    19.82 &    33.72\\
\hline
S\&P~500            &     2.31 &    11.74 &     0.52 &     0.72 &     3.04 &    19.96 &    33.92\\
\hline
\end{tabular}

\caption{Out-of-sample statistics comparing the SPY ETF with a scaled SSD strategy including SPY and options}
\label{table4}
}
\end{table}

Due to tracking S\&P~500 NTR, SPY has a slightly better overall performance when compared to the original S\&P~500 index, with a similar risk profile but higher performance metrics. Using SSD with SPY and options  helped improve all statistics, once again with the exception of volatility. CVaR and MDD show a strong reduction in tail risk. We observe an  increase in volatility, from 19.82\% to 22.67\%. However, as before the increased volatility was concentrated on the right tail (``good volatility''); the addition of options caused the Sharpe ratio to increase only 8.3\% (from 0.60 to 0.65), whilst the Sortino ratio increased 32.1\% (from 0.84 to 1.11).

\begin{figure}[!htb]

\centering
\includegraphics[width=1\textwidth]{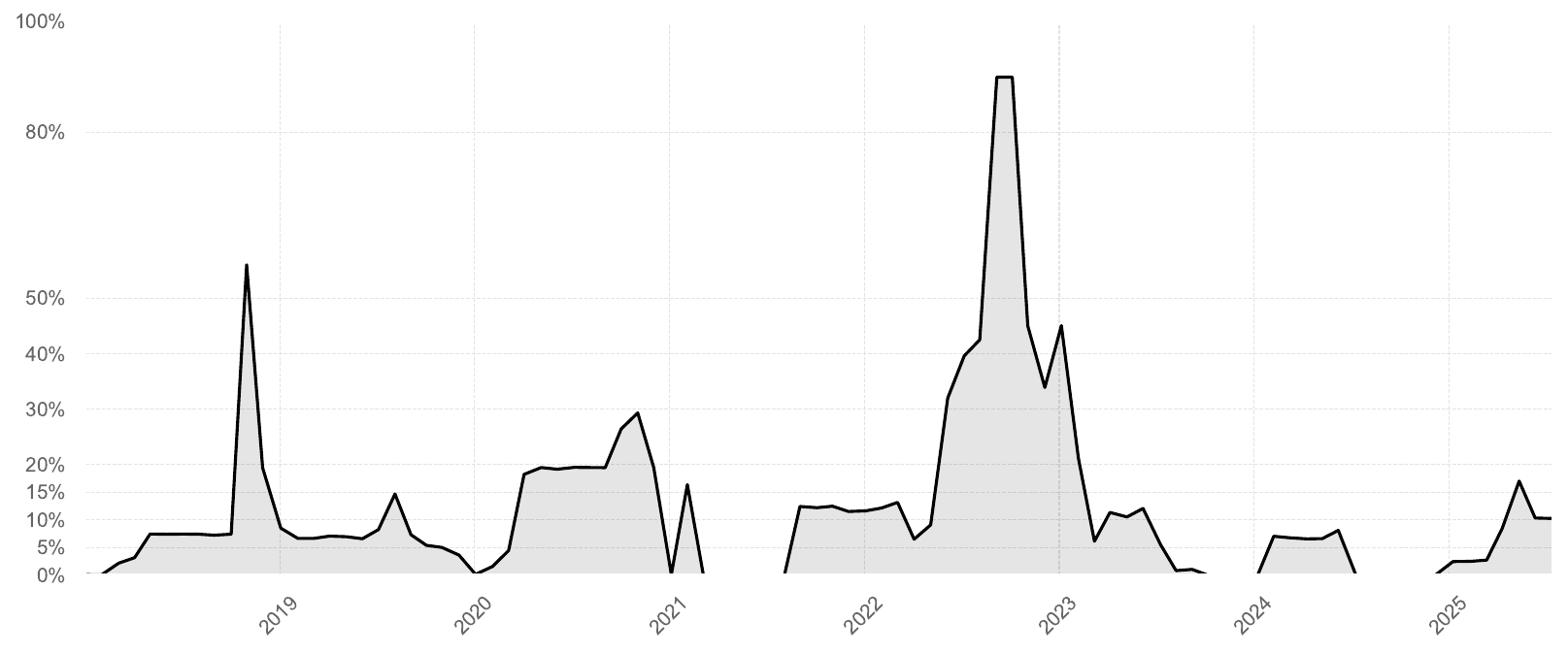}
  \caption{SPY with options: accumulated SSD weights in all 12 option strategies over time}
  \label{fig4}

\end{figure}

From Figure~\ref{fig3} we note moments where a position in options helped offset market drops, such as right before 2019, although it was followed by a subsequent drop when the market recovered (due to negative option strategies performance). Once again, the SSD model was free to choose any exposure in options, and Figure~\ref{fig4} shows that during some periods SSD chose weights in options higher than 50\% (in 2019) and higher than 80\% (in 2022). 

Detailed examination of the results revealed that during the particular period of high option exposure in 2022, the in-sample selected option strategies typically did not ``activate'' out-of-sample, i.e.~there were few 
out-of-sample option trades meaning most out-of-sample investment was just in the risk-free asset, hence the non-volatile performance during a short period in Figure~\ref{fig3}. As noted previously above if necessary exposure limits can easily be added to limit option exposure via use of Equation~(\ref{jebg}).

Table~\ref{table5} compares (scaled) SSD with equities and SSD with SPY, both with options. For four of the six
 performance measures SSD with equities and options  outperforms SSD with SPY and options. For the two measures where SSD with equities and options   performs worse (Sortino and  CVaR) the difference is marginal for Sortino but more significant for CVaR 13.4\% worse (3.21 as against 2.83).

\begin{table}[!ht]

\centering
{\small
\renewcommand{\tabcolsep}{1mm} 
\renewcommand{\arraystretch}{1.4} 
\begin{tabular}{|l|rr|rr|rrr|}
\hline
\multicolumn{1}{|c|}{Strategy} & \multicolumn{1}{c}{FV} & \multicolumn{1}{c|}{CAGR} & \multicolumn{1}{c}{Sharpe} & \multicolumn{1}{c|}{Sortino} & \multicolumn{1}{c}{CVaR} & \multicolumn{1}{c}{Vol} & \multicolumn{1}{c|}{MDD}\\
\hline
SSD: equities and  options & 3.61 & 18.51 & 0.74 & 1.08 & 3.21 & 22.58 & 21.43\\
SSD: SPY and options & 3.12 & 16.25 & 0.65 & 1.11 & 2.83 & 22.67 & 25.57\\
\hline
\end{tabular}

\caption{Out-of-sample statistics comparing SSD with equities and SSD with SPY, both with options}
\label{table5}
}
\end{table}

Overall we believe that the computational results  presented above
illustrate that (for the data considered) introducing option strategies in an enhanced indexation setting 
offers clear benefits in terms of improved out-of-sample performance.

\subsection{Estimated transaction costs}

In the experiments reported above we ignored transaction costs. In this section we consider an ex-post examination of their 
possible impact based on portfolio turnover between consecutive rebalances.
In our work we assume that only the difference between the current and the rebalanced portfolios is traded, so at each rebalance
the fractional turnover is  given by $\sum_{i \in E \cup F \cup S} | \text{(\$ invested in asset $i$ after rebalance)} - \text{(\$ invested in asset $i$ before rebalance)}| /  \text{(\$ invested in asset $i$ before rebalance)}$. 

Let $f$ be the average fractional turnover  over the 91 rebalances in our work. Then if $\alpha$ is the fractional loss in value on the portfolio traded at each rebalance
the value of $\alpha$ that reduces the final value of the portfolio to that of the S\&P~500 at the end of the period is given by
\begin{equation}
(1-\alpha f)^{91} \text{(FV of our approach)} = \text{(FV of the S\&P~500)}
\label{tc}
\end{equation}
So for example for the first case in Table~\ref{table3}, SSD with equities and options,  the average proportion of the portfolio traded was $f=0.4881$, and we have a FV for our approach of 3.61 and a FV for the S\&P~500 of 2.31. From Equation~(\ref{tc}) $\alpha=0.0100$, equivalently 100 basis points  (bps). So if average transaction cost per rebalance is less than this the FV from our approach will exceed the FV of the S\&P~500. Whilst recognising that transaction costs vary according to timing and size of trade
 this value of 100 bps exceeds values commonly quoted for institutional  trading (15-30 bps for a one-way equity trade). Hence  it seems likely from this
 ex-post analysis that even in the presence of transaction costs our approach would result in a final value exceeding that of the S\&P~500.

For SSD without options  in Table~\ref{table3}  $f=0.4856$, giving $\alpha=0.0051$ (51 bps), whilst for SSD with SPY
 and options in Table~\ref{table4}  $f=0.0794$ 
giving $\alpha=0.0247$ (247 bps), based on the FV for SPY of 2.61.

Whilst the above analysis neglects any transaction cost resulting from option strategy trading in each out-of-sample period between 
rebalances detailed examination of the results indicated that this would not significantly alter the values for $\alpha$ seen above. For example as can be seen visually in Figure~\ref{fig2} investment in option strategies only occasionally exceeds 10\% of the portfolio, and is often substantially less. With respect to Figure~\ref{fig4}, where the investment in options exceeds that seen in Figure~\ref{fig2}, note that the $\alpha$ value is much higher, 247 bps as compared with 100 bps.

\section{Discussion}
\label{sec6}

The results shown in Section \ref{sec5} illustrate the  benefits of including option strategies together with equities in terms of enhanced indexation. As mentioned in Section~\ref{sec:ext} however  we believe that option strategies have significant application in a number of portfolio optimisation problems. The basic idea of an option strategy opens the way to including  options of any kind (not just index options) into portfolio optimisation problems.

Clearly, as stated above, we have ignored some practical issues such as transaction costs and  liquidity penalties in our work. This issue is especially important for options, whose bid-ask spread is generally higher than for stocks.  However given that our work covers the time period 2017-2025 it is arguable that ignoring these issues is all that 
could reasonably be done, due to the difficulty of obtaining values for such costs.
However, as discussed above in Section~\ref{sec:eval}, if values for transaction costs and  liquidity penalties can be obtained they can be easily incorporated into option strategy evaluation.


In this work we have priced options using Black-Scholes. It is important to note however that option strategies are \emph{\textbf{independent of the pricing algorithm adopted}}. Rather an option strategy just requires that an option of a particular type with a given lifetime and exercise price can be priced. As such more sophisticated pricing algorithms could be used. \cite{bates22} provides a recent review of some option pricing approaches. In particular we would note here that any financial institutions interested in exploring
option strategies might well have proprietary option pricing algorithms that they wish to use.

To extend the definition of an option strategy given in Section~\ref{sec:str} to deal with American options, which can be exercised at any time up to expiry, the principal change needed is to add a rule specifying the conditions under which the option should be exercised.
Clearly we would need a suitable pricing algorithm for an American option but, as noted in the previous paragraph, option strategies are independent of the pricing algorithm used.

The option strategies used in this paper have demonstrated that relatively simple (but well designed) option strategies have the potential for profit. But the possibilities for development of more sophisticated strategies are endless. 

A particularly promising direction for future research lies in systematically designing targeted option strategies, such as providing downside protection during adverse market conditions, or enhancing performance during favourable market conditions (or both). Moreover it would be naive to believe that a specific option strategy would be equally successful in all markets.  Given the virtually limitless 
number of possible option strategies and use cases, options strategy design will, we believe, constitute a rich area of study (especially for practitioners, but also for academics) in its own right.

\section{Conclusions}
\label{sec7}

In this paper we considered how we can include index options in enhanced indexation. We presented the concept of an option strategy which for a given option is a specified set of rules which detail how the option is to be traded 
depending upon market conditions. An option strategy overcomes the fundamental difficulty that individual options only have a short life and enables us to treat an option strategy as equivalent to an  asset.

We demonstrated the use of option strategies in  enhanced indexation by using an enhanced indexation approach based on 
second-order stochastic dominance. We considered index options for the S\&P~500, using a dataset of daily stock prices from 2017-2025 that has been
manually adjusted to account for index composition. This dataset has been made publicly available for use by future researchers.

Our computational results indicated that introducing option strategies in an 
enhanced indexation setting offers clear benefits in terms of improved out-of-sample performance.

Although we considered option strategies in the context of enhanced indexation we highlighted how they have much wider applicability in terms of portfolio optimisation.

\bibliographystyle{plainnat}
\bibliography{paper}

\linespread{1}
\small \normalsize

\appendix
\section{Cutting plane resolution}
\label{append}
The cutting plane resolution procedure for the portfolio optimisation program Equations~(\ref{jebt4})-(\ref{jebt8}) given above is as follows

First define an initial set of  time periods $\mathcal{J^*}$,  where there is at least one set of time periods of cardinality $s$  in $\mathcal{J^*}$ for all values of $s=1,\ldots,T$. Amend Equation~(\ref{jebt5}) to
\begin{equation}
\mathcal{V}_s \leq \frac{1}{T} \sum_{j \in \mathcal{J}} \sum_{i \in E \cup F \cup S} r_{ij} w_i - \tau_s~~~~\forall \mathcal {J} \in \mathcal{J^*},~|\mathcal{J}| = s,~s=1,\ldots,T
\label{jebt5a}
\end{equation}

\begin{compactenum}
\item Solve the amended optimisation program, optimise Equation~(\ref{jebt4}) subject to Equations~(\ref{jebt5fab})-(\ref{jebt8}),(\ref{jebt5a}).
\item Consider each value of $s$ ($s=1,\ldots,T$) in turn and if in the solution to the amended optimisation program 
\begin{equation}
 \mathcal{V}_s > \frac{1}{T} \text{(sum of the $s$ smallest portfolio returns)} -  \tau_s
\label{jebt9a}
\end{equation}
then add the set of time periods associated with these 
$s$ smallest portfolio returns to $\mathcal{J^*}$. Here the set that is added constitutes a valid cut associated
 with Equation~(\ref{jebt5}) that is violated by the current solution. 
\item If sets have been added to  $\mathcal{J^*}$ go to Step (1), else terminate with the optimal solution to the original linear 
programming  problem,
Equations~(\ref{jebt4})-(\ref{jebt8}).

\end{compactenum}


\end{document}